# Bridging Boundaries: How to Foster Effective Research Collaborations Across Affiliations in the Field of Trust and Safety

Amanda Menking*

Mona Elswah†

David J. Grüning†

Lasse H. Hansen†

Irene Huang†

Julia Kamin†

Catrine Normann†

*Corresponding author: amanda@trustandsafetyfoundation.org

† In alphabetical order

**Acknowledgements.** We'd like to thank Toby Shulruff for early feedback.

# Author Biographies

**Mona Elswah** is a lecturer in digital media governance at the University of Exeter. Prior to this, Mona was a research fellow at the Center for Democracy and Technology (CDT), where she led a project on content moderation for low-resource languages in the Global South. Mona was also a Fellow in Technology and Human Rights at Harvard University's Carr Center, where she examined the algorithmic repression employed by tech companies against Arab users. She has a Ph.D. from the Oxford Internet Institute at the University of Oxford and is currently the co-chair of the Trust and Safety Foundation's Global Majority Research Committee (GMRC).

**David Grüning** is a postdoctoral researcher in digital health and interventions at the Max-Planck Institute and Stanford University. He is also the scientific director of the smartphone apps *one sec* and *Structured*, where he oversees development and testing of new interventions within the apps, and the science board director at the Prosocial Design Network, bridging academic research on digital interventions and the practice of them in real online environments. David completed his Ph.D. in Psychology, specializing in Digital Behavioral Science, at the University of Heidelberg in Germany.

**Lasse Hyldig Hansen** is a behavioral advisor/scientist at the Danish Competition and Consumer Authority (DCCA), where he works on evidence-based policy and consumer protection through statistical and computational methods in behavioral science. He is also a researcher at the Department of Political Science at Aarhus University, focusing on digital regulation and applied behavioral insights. Lasse holds a degree in cognitive science from Aarhus University and serves as a mentor with MIT Critical Data, where he contributes to research and datathons on algorithmic fairness in healthcare.

Professionally trained as a clinical psychologist and data scientist, **Dr. Irene Huang** is a leader capable of leveraging quantitative and qualitative insights to inform business decisions and drive outcomes. As an academic researcher, she is well recognized in the field of suicide prevention, having published over 30 academic articles with approximately 6,000 citations. Dr. Huang currently heads the Wellness and Resiliency department at TaskUs, leading a team with over 200+ coaches and counselors and 10 researchers and data scientists. Dr. Huang specializes in program creation and delivery at scale, user experience research, and intervention research.

**Julia Kamin** is a researcher who works with organizations that leverage social science to foster social cohesion on and off line. Currently Julia is Managing Director at Prosocial Design Network, a non-profit that connects research to practice to promote healthy online spaces. She also works as a research consultant for Civic Health Project on programs related to measuring social cohesion and bridging divides in the US. Julia completed her Ph.D. work in Political Science at the University of Michigan.

**Amanda Menking** is a qualitative researcher with experience in human-computer interaction (HCI) and information science. She joined the Trust and Safety Foundation (TSF) after spending almost a decade in academia, researching bias, knowledge production, and safety in online communities (e.g., Wikipedia, Reddit). Amanda is also a veteran of the Seattle tech scene, having worked for two start-ups and as a vendor at Microsoft prior to completing her Ph.D. in Information Science at the University of Washington.

**Catrine Normann** holds a Ph.D. in Behavioural and Experimental Economics from the University of Copenhagen. Her research spans academic, governmental, and industry settings, with a strong focus on experimental approaches. She currently serves as Senior Behavioural Insights Advisor at the Danish Competition and Consumer Authority (DCCA), where she supports public policy design aimed at improving consumer protection and competition. Catrine also holds a specialist role in Experimental and Research Methods, advising governmental institutions on methodology and design both nationally and internationally.

## Abstract

As the field of Trust and Safety in digital spaces continues to grow, it has become increasingly necessary - but also increasingly complex - to collaborate on research across the academic, industry, governmental and non-governmental sectors. This paper examines how cross-affiliation research partnerships can be structured to overcome misaligned incentives, timelines and constraints while delivering on the unique strengths of each stakeholder. Drawing on our own experience of cross-sector collaboration, we define the main types of affiliation and highlight the common differences in research priorities, operational pressures and evaluation metrics across sectors. We then propose a practical, step-by-step framework for initiating and managing effective collaborations, including strategies for building trust, aligning goals, and distributing roles. We emphasize the critical yet often invisible work of articulation and argue that cross-sector partnerships are essential for developing more ethical, equitable and impactful research in trust and safety. Ultimately, we advocate collaborative models that prioritize inclusivity, transparency and real-world relevance in order to meet the interdisciplinary demands of this emerging field.

*Keywords:* cross-sector collaboration, trust and safety, interdisciplinary, digital space, collaboration framework

# Introduction

Imagine a group of researchers coming together to determine how to best measure the psychological wellbeing of content moderators. Now, imagine this group consists of: two clinical psychologists working for a business process outsourcer (BPO); a tenured professor; a PhD candidate; a user researcher working for a tech company; a data scientist working for a BPO; and a researcher employed by a non-profit organization. How might these individuals collaborate in a way that leverages and negotiates the responsibilities, resources, and constraints of their respective roles—with or without support from their employers?

This paper explores these dynamics: research collaborations across different affiliations, specifically in the field of Trust and Safety. Given the rapid evolution of this field, it's essential to build strong multistakeholder approaches through which researchers from different backgrounds can collaborate, advancing scientific knowledge and addressing challenges in emerging areas. However, researchers from different affiliations often encounter obstacles in aligning their priorities, incentives, constraints, and methodologies when working together. Additionally, their affiliations may be perceived as competing, with a tendency for their research outputs to be viewed as divergent rather than complementary.

To guide future research and inform collaborative practices, this paper—written by authors from different sectors—outlines several strategies aimed at fostering more integrated, mutually beneficial partnerships. We advocate for empowering cross-sector collaborations that leverage the strengths of each affiliation. We offer a practical framework of collaboration that prioritizes inclusivity and impactful outcomes. We begin by defining key terms and identifying affiliations; we then discuss key differences and similarities between affiliations. In conclusion, we propose a collaborative approach for future partnerships to advance scientific knowledge in the field of Trust and Safety.

## Definitions

While the range of collaborations between researchers could include drafting literature reviews, authoring research-informed design guides, or writing policy articles, for the purposes of this article we're specifically interested in collaborations focused on **original research**. We define original research as intentional, systematic, and careful consideration of a given question, problem, or phenomenon, which can make use of an array of approaches and methodologies. A researcher, by our definition, is a person who has been professionally trained to use scientific methodologies to engage in the creation of new knowledge, products (for example, R&D), processes, methods, systems and/or theories. We are likewise focused on original research related to **Trust and Safety**, "the study of how people abuse the internet to cause real human harm, often using products the way they are designed to work" (Cryst et al., 2023).

Trust and Safety related research is rapidly evolving across all sectors. However, cross-disciplinary collaborations are limited, and researchers from different backgrounds often don't work together, restricting potential impact. This isolation can result in academics having less access to industry data, industry researchers lacking robust theoretical insights, and both governmental and nongovernmental researchers facing challenges in data access and the implementation of their recommendations. We encourage collaboration across affiliations to foster a more inclusive and effective research environment.

## Current Context

Today, more and more people are collaborating across affiliations when it comes to Trust and Safety related research. Organizations explicitly aimed at connecting and supporting researchers and practitioners facilitate many of these efforts.[1] Relationships also arise between researchers by virtue of the fluidity with which many move between affiliations, in particular with graduates from multiple disciplines (i.e., CSS, IS, HCI, Communications, Social Psychology, Political Science) working within industry. Of note, there are several examples of *ad-hoc* collaborations between research institutions and platforms that have resulted in published studies (for example, see Clegg & Nayak, 2020; Kim et al., 2022; Katsaros et al., 2022; Horta Ribeiro et al., 2025).

Others in this space have published related work with slightly more narrow foci about how to best collaborate and work across stakeholder groups. For example, using a Team Science lens, scholars investigated how 17 early and mid career human-computer interaction (HCI) faculty from 16 different universities in the U.S. share projects with health researchers, noting challenges and identifying suggested best practices for cross-disciplinary collaboration (Agapie, Haldar & Poblete, 2022). Similarly, others (de Gois Marques et al., 2023; Garousi, Petersen & Ozkan, 2016; Marijan & Sen, 2020) have studied industry-academia collaborations in software development and interdisciplinary intersections in Trust and Safety (Huang, Massullo & McLoughlin, 2023). This paper complements previous work by broadening the conversation to include researchers affiliated with both nongovernmental and governmental organizations.

# Part One

## Types of Affiliations

**Academia.** Academics from a range of disciplines investigate the social, technical and ethical dimensions of digital tools and platforms. The institutions involved focus on theoretical as well as applied research funded by grants or government programmes. Historically, academic researchers are more focused on theoretical conceptions of issues, questions, and solutions (i.e., developing a theory of cognitive inoculation to counter misinformation online). However, academic research has begun to move into more applied areas as well, with practical impact in mind.

Academic researchers make significant contributions to Trust and Safety related research through their expertise in a range of disciplines. For example, social scientists, such as those in Communications, Political Science, and Social Psychology, most often study the societal impact of digital platforms and tools. For another example, computer scientists in academia focus on the development of foundational technologies (e.g., algorithms, artificial intelligence, cybersecurity measures) that underpin digital systems.

---

[1] These organizations include the Trust and Safety Foundation (TSF), Prosocial Design Network (PDN), All Tech is Human (ATIH), Integrity Institute, and the Coalition of Independent Tech Researchers (CITR). Events hosted by these organizations as well the Trust & Safety Professional Association (TSPA) and several applied academic disciplines are also focal points for connection and collaboration; to name several prominent events: Trust & Safety Research Conference; Conference on Computer-Supported Cooperative Work & Social Computing (CSCW); Conference on Human Factors in Computing Systems (CHI); Web Science Conference (WebSci); and International AAAI Conference on Web and Social Media (ICWSM).

**Industry.** Technology companies, from start-ups to large corporations to vendors, play a key role in Trust and Safety research, especially for applied areas. Industry researchers often work on application-oriented projects, such as improving user experience, developing algorithms for user engagement and content distribution, or designing new products suited to needs voiced by the community. For example, a tech company might study how changes to the design of its platform affect user engagement or safety.

**Governmental organizations.** Governmental and intergovernmental (i.e., United Nations, World Economic Forum) institutions play an increasingly proactive role in synthesizing and integrating research insights into public policy and regulation, particularly by drawing on behavioral science and evidence-based methodologies. Government-affiliated researchers thus occupy a unique position: they are tasked with applying rigorous methods to real-world challenges, often under political and legal constraints, while also acting as bridges between sectors. As digital technologies continue to evolve, governments are likely to expand these collaborative efforts, particularly where public interest, innovation, and accountability intersect.

**Nongovernmental organizations.** For the purposes of this article, we have elected to use the broader category of nongovernmental organizations (NGOs) to include a range of types of affiliations (e.g., non-profit, civil society, think tank) often distinct from academia, industry, and government. However, there are variations among researchers who work for different types of NGOs, and reporting lines may be blurred (for example, some think tanks report to governmental bodies). Additionally, not all researchers affiliated with NGOs conduct original research.

*Non-profit organizations.* Non-profit organizations often focus on issues such as digital rights, ethical technology development, and access to technology. NGOs aim to inform policy or raise awareness, but several focus on building prosocial tools. For example, a non-profit might research the impact of digital surveillance on marginalized communities and propose regulatory frameworks to address it. Non-profit organizations may be funded by individual donors or other non-profits; sometimes they receive funding from industry and/or governments.Researchers affiliated with non-profit organizations often focus on conducting original research that advances advocacy, education, and policy development in digital technology; however, they may also engage in research that is more advocacy-motivated.

*Civil society.* Civil society includes independent researchers, consultants, and community-focused innovators who work outside traditional institutional frameworks. These individuals often engage in grassroots projects or cross-sector collaborations, addressing issues such as technology access or the ethical use of AI in underserved populations.

*Think tanks*. Like non-profit organizations, think tanks may be funded by individual donors, other non-profits (e.g., foundations), and/or governments. Some think tanks are even situated within governmental or military organizations.

Exclusions

We have excluded some affiliations here (e.g., such as media, press, marketing, and public intellectuals) due to their distinct objectives and because their work more often serves to disseminate rather than generate empirical evidence and insights.

# Part Two

## Differences and Similarities

**Priorities, Incentives, and Impact Metrics.** Although many researchers receive similar training—especially if they come from the same discipline—where they end up working greatly impacts *why and how* they conduct research. Researchers work within systems. Recognizing and acknowledging structural realities can help collaborators across affiliations to better understand and negotiate with each other during shared projects.

**Academia.** Academic researchers strive to make intellectual contributions to their disciplines or subdisciplines; intellectual contributions are original, evidence rigor, and have the potential to advance a field of study. Academia structurally incentivizes intellectual contribution through a limited yet compelling range of rewards (i.e., tenure) and assesses intellectual contribution using metrics like publication records (e.g., number of publications, citations, impact factor, h-index), invitations to join prestigious societies (for example, National Academy of Sciences), and grants received. In some fields and disciplines, the commercial application of research is also used as a metric to assess impact (i.e., new medical discoveries that lead to drugs, devices, treatments, or protocols). While many grant applications and even tenure discussions may include questions about "broader impact," career trajectories are largely determined by impact *within* the academy.

Historically, academic research has been both narrow (versus broad) and siloed (versus interdisciplinary). Academic researchers often "go deep," conducting intensive studies about a given topic, phenomenon, or population over several years; in this way, they establish their expertise and cement their reputation within their field. In recent decades, however, more interdisciplinary programs have been established, creating new fields that promote working across established, siloed disciplines (i.e., Information Science). Regardless, even in more interdisciplinary fields, intellectual contribution is the primary incentive and motivation for academic research.

**Industry.** Unlike academic researchers, industry researchers are generally not incentivized to make intellectual contributions to a field or discipline but rather to conduct research that aligns with and can further the aims of the business. (There are, of course, always exceptions to this rule, including labs such as Microsoft Research Cambridge and units such as Jigsaw, at which researchers may do both—often in collaboration with researchers affiliated with academia or nongovernmental organizations.)

The tech industry structurally incentivizes research that aligns with business interests by financially rewarding researchers for engaging in work that furthers their employer's business objectives. For example, companies may require researchers to meet key performance indicators (KPIs) for themselves and/or for their teams, track the number of projects or studies per quarter, and evaluate how research has impacted a product and/or revenue (e.g., new features, user growth, user retention rates, checkout rates) to receive promotions, raises, and/or stock. Researchers employed by private companies are rewarded when their employers receive funding (in the case of start-ups), profit, and/or are acquired by or merged with larger companies.

**Governmental organizations.** Researchers in governmental organizations work under a public interest mandate, with priorities shaped by institutional responsibilities, legal frameworks, and societal concerns. While methodological rigor is essential across all research domains, the standards in governmental contexts are oriented toward public accountability.

Governmental researchers are expected to produce policy-relevant evidence that is defensible under public scrutiny. Their work is assessed less by theoretical novelty or citation metrics and more by its neutrality, methodological robustness, and ability to inform actionable decisions: particularly whether legislation, regulation, or intervention is needed, likely to be effective, and enforceable within existing systems.As a result, speculative claims and polarizing framings are generally avoided.In practice, this often requires a value-sensitive framing of research questions. Instead of asking, for example, whether screen time is "bad" they start from shared policy goals—such as ensuring that digital environments do not undermine youth mental health—and work to evaluate interventions in terms of feasibility, and legal justification.

Incentives in this context are tied to usability in governance. A project is considered successful if it provides credible evidence that supports, refines or challenges policy directions. Impact is measured not by academic indexes or product performance, but by policy uptake by institutions or citizens, or by improved enforcement efforts. This could be whether the research shapes regulatory language or underpins a legislative proposal, or the implementary success of policy initiatives through citizen uptake, for instance.

**Nongovernmental organizations.** While they share similarities with industry researchers in terms of timelines, ethics, and procedures, NGO researchers aim to inform the public and influence policy, with their work designed to be publicly accessible and impactful. For NGO researchers to achieve impactful research outputs, media attention is needed and researchers actively work hard to attract media coverage. Additionally, another success metric would be their ability to attract external grants and support for their work.

**Constraints, Timelines, Audiences, and Deliverables.** Researchers affiliated with different kinds of organizations must work within different systems of constraints, all with differing timelines, audiences, and deliverables. Our goal in outlining these differences is to enable collaborators from different stakeholder groups to more effectively identify potential challenges and address them early on in the project. Awareness of these challenges may also guide researchers in seeking out complementary partnerships, which we discuss in more detail in Part Three.

**Academia.** Despite enjoying a great deal of intellectual freedom, academic researchers face many constraints in their work. They often lack direct access to data and communities outside of academia and, therefore, must invest a great deal of time and resources in establishing partnerships and/or gaining access to data as well as establishing relationships of trust with communities to which they do not belong. Additionally, academic researchers may not be able to conduct studies at scale in the same way that industry researchers can simply because they lack access to the same quantity of data (for example, data collected by social media platforms understand Terms of Service) and/or lack the budget to recruit and process the same number of participants.

In terms of securing funding to conduct research, researchers affiliated with academia face high competition for limited resources (for example, National Science Foundation award rates are generally lower than 30%). Although some universities provide tenure track faculty with "start-up funds" when they are first hired, the majority of researchers affiliated with academia spend a non-trivial amount of time trying to secure funding (i.e., writing grant proposals).

Most researchers in academia face time constraints as they are also teaching professors responsible for creating and delivering courses as well as advising students; additionally, tenure track faculty are expected to engage in service (i.e., peer reviews, committee membership).

In terms of timelines, academic research often spans months, years, and even decades. Many of the deadlines in academic research are driven by the research questions themselves (for example, Harvard has been running a study of adult development for more than 80 years due to the nature of the initial research questions) and by grant cycles and/or venue and publication deadlines, both of which generally follow annual and semi-annual patterns. Because peer review is central to the academic publication process, it can take months to years (though the advent of pre-prints is easing this), and, in general, applied research is less likely to be published.

When it comes to academic research, other scholars in the author's discipline are the primary audience; that is, academics write for other academics. Secondary audiences may include academics from other, closely related disciplines, and tertiary audiences may include those outside of academia (i.e., governments, lay people). Because academics primarily write for other academics, they generally focus on publications in peer-reviewed journals or prestigious conference proceedings, prioritizing conferences directly related to their discipline (especially if they have limited funds for travel).

**Industry.** Researchers affiliated with industry face their own sets of constraints. Because they are employed by private companies, they generally work under strict legal conditions, including non-disclosure agreements (NDAs), and are discouraged from publishing unfavorable findings that may impact their employer's reputation or profits. Industry researchers can also be averse to sharing information, not only because of NDAs, but also due to concerns about competitive intelligence. Additionally, individual researchers may not have the agency to prioritize certain questions or topics as research agendas are generally set using a top-down approach and timelines are determined by product releases and/or business objectives. Even when an industry researcher or their team is able to conduct research and present findings, the findings may be disregarded if they don't align with other priorities.

As noted above, in terms of timelines, industry researchers must adapt to the needs of their company and to other teams; however, depending on the research and the environment (e.g., company stage, company culture, internal processes and methods), research could be conducted over weeks or months. For example, user researchers working within a company that relies on sprints may move very quickly.

In general, researchers affiliated with industry conduct research to present to their immediate teams and managers and to cross-functional stakeholders within their company. Secondary audiences may include partners or collaborators (in formal programs under NDAs). Because of this, industry researchers are likely to spend the majority of their time creating deliverables for the primary and secondary audiences, and that generally looks like decks or presentations for internal meetings. At times, industry researchers will also present at industry specific conferences (i.e., IXDA, Marketplace Risk, TrustCon) or at conferences aimed at other researchers in industry (for example, EPIC). For researchers who are affiliated with industry but who work at labs or have roles that allow them to collaborate with academics, they may also co-author papers for peer-reviewed journals and/or participate in academic conferences (for example, CSCW, CHI, and DIS). The majority of industry researchers, however, spend their time focused on research for internal stakeholders that result in concrete changes (i.e., new features, prototypes, improved processes).

**Governmental organizations.** Governmental organizations operate under conditions shaped by statutory responsibilities and democratic processes, which set specific boundaries on how research is framed, timed, and communicated. Here, original research is scoped to support policy-relevant questions and expected to reflect values that are already broadly accepted or legislatively grounded.

The timelines governing governmental research are shaped by political calendars, regulatory windows, and administrative planning cycles. While some projects are long-term and strategic in nature, most operate under timelines within the calendar year. Before a project begins, its timing is typically aligned with current public or political agendas, which are in turn shaped by societal developments and emerging issues.

Governmental research also serves multiple, often simultaneous audiences. These include policy advisers, legislative staff, legal experts, agency heads, and the broader public. Each audience has distinct expectations. For example, ministers may require short, actionable takeaways, while legal departments may need precise definitions, statutory reasoning, and case comparisons. Public-facing documents must meet high standards of clarity, neutrality, and accessibility.

Deliverables are equally varied: internal memos, slide decks, implementation advice, or public reports. In many cases, deliverables are not designed for wide distribution, but for targeted institutional use. Even when publishing is possible, it typically involves a clearance process to ensure alignment with government positions. External dissemination may be constrained by political timing or sensitivities around messaging. These considerations can affect everything from framing to authorship decisions in cross-sector collaborations.

To work effectively with governmental researchers, collaborators must recognize that research quality alone is not sufficient. Outputs must also be usable within administrative workflows, legally sound, and publicly defensible. Aligning expectations on pace, positioning, and review processes is essential to ensure that joint work can be implemented without compromising institutional mandates.

Finally, conducting governmental research sometimes depends solely on collaborative efforts to be possible. Specific talent, data access, and research capacity might not be possible or sensible to acquire in-house, and thus depend on collaborative inputs from other governmental institutions, industry, or academia, or be out-sourced entirely.

**Nongovernmental organizations.** Like academic researchers, researchers affiliated with NGOs may struggle to access data and secure funding. They are also limited by the mission of the organization and by board pressures—and they may be perceived as producing less credible, less rigorous work products because they don't go through the same review processes (e.g., Institutional Review Board, ethics committee, peer reviews). Unlike researchers affiliated with governments, nongovernmental researchers may be disincentivized from pursuing projects that center broad, multistakeholder perspectives because many NGOs benefit from both media attention and financial donations when they serve as advocates (i.e., leading a boycott).

Similar to industry researchers, researchers affiliated with NGOs are expected to produce multiple reports annually to meet funding and internal deadlines. Since NGOs rely heavily on external funding, they must fulfill grant expectations, which often places pressure on researchers to publish more work compared to their counterparts in academia. This dynamic can lead to a faster-paced research environment, where the emphasis is on meeting deadlines rather than pursuing in-depth, long-term studies.

A Note About Ethical Review Boards

If academic researchers plan to involve human participants or human data in their studies, they are required to submit their research plans to their university's Institutional Review Board (IRB) before they begin recruitment. While some companies have internal review boards similar to IRBs, many do not. Researchers working for NGOs generally do not have access to internal review boards and so they must either forgo the process or rely on commercial IRB companies whose services are often cost-prohibitive.

# Part Three

## Unique Strengths: *Where to look?*

In a field as complex and fast-moving as Trust and Safety, no single type of affiliation can address all relevant questions or produce universally applicable solutions. To foster meaningful and effective cross-affiliation collaborations, it's essential to understand the distinct strengths each affiliation brings to the table. Recognizing and leveraging these unique assets is key to designing partnerships that are both complementary and high-impact.

**Academia.** Academic institutions offer high levels of credibility, transparency, and theoretical rigor. Researchers in academia are trained to conduct methodologically sound studies and are often deeply embedded in their disciplinary research literature. Their work is guided by peer review and public dissemination of results, trying to ensure that academic research is accountable, replicable, and can be openly discussed.

In addition, academia is uniquely positioned to advance fundamental knowledge and contribute conceptual frameworks that shape how emerging technologies are understood. Importantly, academic researchers often conduct research on longer time horizons. Academic research also provides a relatively protected space to explore controversial or high-risk topics without the same commercial or political constraints that other affiliations may face.

Further, academic researchers, through teaching responsibilities and duties, also play an important role in training the next generation in fields such as technologists, data analysts, and policymakers, thereby shaping the intellectual and ethical foundations of the Trust and Safety field as a whole.

**Industry.** Industry-based researchers have access to unparalleled speed, scale, and real-world implementation. They often operate within agile environments where new features and tools can be developed and tested quickly and instantly deployed afterwards. This ability to move fast enables industry researchers to iterate rapidly, naturally test ideas in real-world settings, and generate data at a scale rarely possible in academic or nonprofit contexts.

Perhaps most notably, industry can directly affect (as in change or advance) the digital environments that billions of people interact with daily. A single design tweak—such as changing the layout of a news feed or adding a friction point in a user flow—can produce immediate and measurable effects across entire online populations. When industry researchers apply their findings, the real-world impact is often substantial and swift.

Industry researchers also tend to have extensive access to already existing behavioral data, providing insight into trends and interactions that cannot be observed in smaller scale settings. This makes industry research an ideal context for understanding usage patterns in detail, identifying emerging harms, and prototyping solutions that are grounded in actual user behavior.

When partnered with academic or (non)governmental collaborators, industry teams can serve as real-time laboratories for testing interventions, scaling specific experimental efforts, and exploring edge-case phenomena (e.g., reactions to an intervention by existing user subgroups). However, collaborators must also be mindful of the commercial pressures that may limit what can be studied or published.

**Governmental organizations.** Governmental organizations bring a distinct set of contributions to collaborations in responsible tech research. Rather than advancing new theoretical paradigms, governmental researchers focus on whether policies and interventions can be made to work in practice—legally, operationally, and in terms of enforcement. Their role often involves translating research findings into options that are compatible with existing regulatory frameworks. In this way, they serve as a link between original research and real-world decision-making, helping to assess whether policy action is possible, and if so, what form it might take.

In terms of capacity, government institutions can, when needed, dedicate concentrated resources to high-priority topics. Researchers may be given extended timelines and institutional backing to investigate a particular issue in depth, especially when the topic aligns with pressing societal needs. There is no parallel pressure to publish a certain number of articles or teach, which can enable teams to focus their full attention on one applied research task.

Finally, governmental organizations often have a built-in platform for communication. Their position in the public sphere means that research findings—when ready and cleared—can be shared at scale, whether through public reports or press briefings. This capacity helps ensure key insights reach both policymakers and the broader public in ways that are coordinated and institutionally anchored.

**Nongovernmental organizations.** NGOs play a distinct and increasingly vital role in Trust and Safety related research. They often serve as bridges between academic theory, industry application, and the public interest. One of their key strengths lies in their mission-driven focus. NGO researchers are typically guided by and have the possibility to pay close attention to principles such as equity, justice, transparency, and accountability. Their work is often motivated by a desire to call attention to underexplored harms, advocate for marginalized communities, and shape ethical standards for the technology sector.

In contrast to industry and more in line with academic research, NGOs are less bound to profit motives or internal product development, giving them the opportunity to pursue pioneering ideas that challenge dominant paradigms and contrast commercial interests. This often leads to first-mover efforts in areas such as digital rights, content moderation ethics, and algorithmic accountability, which then lead to these topics later receiving attention from industry or governmental bodies.

NGOs are focused on producing public-facing outputs—ranging from reports and op-eds to briefings and workshops—that help translate complex research into accessible formats for policymakers, journalists, and civil society at large. Their unique freedom to combine research with advocacy allows them to generate issue-focused insights and impact from these insights in a way that academic publications or internal industry memos often can only do in restricted ways. NGOs are also uniquely positioned to amplify marginalized voices, particularly when they maintain close ties to selective, affected communities. Their research often centers those communities not just as subjects of study, but as collaborators, advisors, or co-creators—as a result, strengthening both the legitimacy of their research and the impact of the resulting implications.

# Initiating collaborations: *What to consider when launching effective cross-affiliation partnerships*

Collaborations across academia, industry, and (non)governmental institutions can be uniquely powerful—but also uniquely complex. As outlined above, different goals, incentives, timelines, and operational cultures can create friction if not navigated intentionally. The following section attempts to offer practical guide posts for initiating and managing collaborations in a structured manner, so that different parties' goals are strategically aligned, and the collaborative effort is successful. To provide actionable recommendations in this section, we present the initiation and management of such collaborations, respectively, in successive steps of the process and what can be done in each of these steps.

**Know when to collaborate.** An important decision point to start at is the question of if cross-sector collaboration is meaningful in one's specific case. Certain research questions or testing goals strongly benefit from it, however, not every project requires cross-sector collaboration. One central reason for such collaborations is if the different sectors promise complimentary expertise, experience, and knowledge for the various elements of the planned project. That is, if such competencies seem to be incomplete or even biased if developed in isolation of one sector of research—for instance, a theoretical model might lack grounding in real-world use cases, or an industry study might overlook structural or community-level impacts—cross-sector collaborations are a meaningful method to address this shortcoming.

Further, such cross-sectoral collaborations are meaningful if the starting sector has concrete practical constraints to conducting the intended research. Two prominent practical restriction cases: one sector can simply not conduct the research (i.e., legal, ethical, or reputational concerns may prevent a company from running a study themselves); one sector lacks certain tools or resources to rigorously conduct the research (i.e., NGOs might not have the funding resources for large scale testing).

In short, if the research needs to be trusted, applied broadly, or cover blind spots from a single perspective, cross-sector collaboration should be seriously considered.

**Identify the right partners.** Once one has decided on a cross-sector collaboration, the next step concerns figuring out which collaborative partners are most suitable for the project. When identifying partners, we suggest specifically focusing on collaborators who bring complementary skillsets as well as complimentary resource access, while having a shared commitment to the goals of the project. Regarding the complementary skillset, we suggest to focus on collaborators' domain expertise (e.g., in misinformation, teen well-being, or algorithmic fairness) and their methodological fit (e.g., quantitative vs. qualitative, experimental vs. ethnographic). Collaborators should either act as a second perspective on some specific aspect of the research (e.g., providing a second opinion on how a misinformation intervention should be constructed or how to successfully conduct a pre-survey) or advance existing expertise (e.g., being the expert for how to intervene with misinformation effectively and how to correctly survey questions).

Besides the skillset and resource considerations, collaborators should also align in their research mission, that is, which goals they attempt to achieve with the research project. A diagnostic indication for this can be a potential collaborators' track record to identify central motives of their research by past behavior, such as where they have published, what tools they have deployed, what projects they have managed, and what they have advocated for in the digital technology space.

**Build trust early.** While not mandatory, it can be advantageous to start processes with potential collaborators early to build trust in one's own clarity of intentions. Such trust is often started with lightweight outreach to the potential collaborators—a short message or in-person introduction. This outreach should focus on concretely communicating the specific topic interested to collaborate on, provide an (at least broad) estimation of the capacity needed from the person, and discuss transparently what the goals of the project are.

If potential collaborators note interest, a more extensive exchange usually makes sense to move toward a research agreement or memorandum of understanding (MoU). In this phase of understanding and alignment, we suggest to focus on some concrete pillars of the collaboration:

- Roles and responsibilities of different collaborators;
- Timeline and milestones that match with people's personal timelines;
- Publication and communication expectations of the project and its results;
- Authorship and (if applicable) data ownership;
- Contingency planning (e.g., what if funding falls through or a key partner exits?), especially in more complex team constellations.

We also suggest treating the MoU as a living document, i.e., not just a legal and structural boilerplate in the beginning of collaboration initiations, and revisiting it when major project pivots occur. One should also consider signing a Non-Disclosure Agreement (NDA) during the communication phase with potential collaborators if sensitive or proprietary information will be shared. This is particularly important when discussing internal datasets, pre-release tools, or reputationally sensitive research topics.

**Align on study design and workflow.** Once trust is established and the project structure of the collaboration is decided, one should shift to planning the project execution in detail. Here, we argue that co-designing the research plan and process is critical. The central topics to align on are:

- Study goals and hypotheses;
- Target populations and recruitment method(s);
- Tools and interventions (if any) to test;
- Data analysis plan;
- Ethics and IRB approval requirements.

For all of these aspects, it is not only important to align on content (e.g., what are the central study goals) but also who is responsible for different aspects (e.g., one collaborator being especially suited to organize the IRB approval for the planned study). Important for these responsibilities is to also align the team on not just timelines for specific goals and achievements but to also transparently communicate and take into consideration collaborators' individual working rhythms, ranging from simple questions like "Do we meet weekly or monthly?" to more complex aspects of who can work on which aspects at which time (be it first drafts or feedback).

**Communicate and document.** As an overarching suggestion, when initiating a collaboration, it is advisable to document all decisions in the initiation process to refer back to. This also includes providing recaps of meetings and discussions as well as a shared action space that fits the research's requirements to efficiently channel information exchange (e.g., Google Docs, Notion, GitHub). These basic practices are almost mandatory to reduce confusion among team members and keep multi-institutional goals aligned.

Importantly, especially in long-term and multi-phase collaborations, communication tends to degrade with time unless proactively maintained. To mitigate these developments, decide on clear check-ins and review points in the research timeline. Also make space to critically engage with concerns within the team and recalibrate if needed, and explicitly celebrate progress to keep momentum.

## Managing collaborations: *Ground rules for effective coordination across sectors*

Once a collaboration has been initiated, it takes active stewardship to ensure it succeeds. Organizing a cross-sector project is not just about moving pieces; it's especially challenging because it includes managing people, expectations, and constraints across potentially very different working cultures. Here, we attempt to offer a practical framework of the cross-sector collaboration type that also includes inclusivity, ethical considerations, evidence-based actions, and impactful outcomes. These suggestions mostly outline how to concretely plan many of the aspects decided on in the initiation phase of the collaboration (e.g., different collaborators' responsibilities and timelines).

**Acknowledge and navigate differences.** Arguably the most important aspect of cross-sector collaborations is to not assume that collaborators in a different sector think or work the same way about the research topic as oneself. Notably, affiliations like "academic," "industry," or "civil society" can be misleadingly broad, and the way individuals approach research can vary significantly—even within the same sector.

Rather than glossing over these differences, it is important to transparently surface them early and discuss their flexibility and adaptability for the project. Doing so allows collaborators to set realistic expectations, avoid misunderstandings, and design a workflow that plays to everyone's strengths. Importantly, and as already indicated above, openly acknowledged and respected, these differences can become assets that enrich the collaboration.

**Align personal and institutional goals.** Each person involved in the project naturally has their own objectives and weighs the outcomes of the project differently. For instance, one collaborator might focus most on the resulting project publication, another on the internal report, and yet another on a deployable intervention that results from the project. These goal differences have to be made explicit to avoid temporal and interpersonal friction. The authors' suggestion is to initiate a conversation around this topic about what each person and their respective institution expects from the collaboration. To do so, ask as concrete questions as possible, such as: "What would a successful project be measured by for each person?" or "What does each collaborator expect to learn, influence, or build through this project?"

**Be clear on scope, timelines, and practical resources**. Effective collaboration across sectors requires early alignment on project scope, timelines, and resource commitments. Discrepancies in institutional pacing can be challenging even for the most seasoned collaborators. Additionally, administrative processes such as contract negotiation, data-sharing agreements, and regulatory approvals often entail significant delays, regardless of sector.

To mitigate these risks, collaborators should begin by establishing a shared understanding of key project parameters. This includes identifying a feasible project initiation date and mapping critical decision points in relation to publication deadlines, review processes, and organizational reporting cycles. Attention should also be given to any internal or external sign-off requirements that may impact progress.

Beyond temporal considerations, scope and feasibility must be clearly discussed and planned. This involves project aspects such as defining the minimum viable sample size and the central variables to collect necessary to arrive at, specifying access needs (e.g., to data, analytical tools, or participant populations), and outlining contingency plans in the event of unforeseen constraints. These elements should be synthesized into a detailed project timeline that includes milestone checkpoints, assigned responsibilities, and documented fallback strategies.

**Assigning roles and responsibilities.** Every collaborator brings unique value, so one meaningful approach to dividing responsibilities in the project effectively is to do so by the profile of skills and access to specific resources that a collaborator brings to the table. Important considerations for collaborator profiles are, for example:
- Their expertise and experience in the topic or a subtopic of the intended research project;
- Hard skills, such as the technical implementation of a design or analysis of data;
- Access to practical resources (e.g., a sample of participants, existing data, analysis tools, or financial support to fund the project), and;
- Access to personnel and social resources such as institutional services (e.g., a grant team at a university) or connection to certain online groups (e.g., Wikipedians or reddit moderators).

As outlined in the collaboration initiation phase, a transparent documentation of the assigned roles holds collaborators accountable for the project's success and avoids misunderstandings such as the duplication of work or missing specific tasks. Early adoption of the role transparency can also serve as a source of motivation, as it affirms each collaborator's contribution to the project's overall success and provides them with adequate recognition for it.

One specifically sensitive issue that early role assignment helps to prevent is that of authorship. Ambiguity around authorship frequently causes tension in research collaborations, yet it is also one of the most easily preventable issues. This is especially pertinent in cross-sector partnerships, where assumptions based on sector-specific norms can lead to misunderstandings. Even within a single sector, institutions often differ in how they recognize and reward contributions. Therefore, it is crucial to openly discuss authorship expectations at the beginning of a collaboration and document key decision-making processes transparently and inclusively.

## Conclusion

      Returning to our opening scenario: How might two clinical psychologists and a data scientist working for a BPO, a tenured professor, a PhD candidate, a user researcher working for a tech company, and a researcher employed by an NGO come together to determine how to best measure the psychological wellbeing of content moderators?

      If this group follows the recommendations outlined above, they might begin by discussing how they can leverage their unique strengths, resources, and perspectives—and whether it makes sense for them to collaborate at all. Then they might openly address how they'll work together amidst constraints and conflicting priorities, agreeing on roles, responsibilities, timelines, audiences, deliverables, and authorship before addressing research design.

      As Schmidt & Bannon argued in 1992, at the heart of all cooperative projects is *articulation work*—"a kind of supra-type of work in any division of labor, done by the various actors" (Strauss, 1985). Though it is often invisible, leaving no trace in the final products of successful collaborations, it is essential to integrated, mutually beneficial partnerships, especially in the field of Trust and Safety. We can't expect better digital technologies without better teamwork—cross-affiliation research collaboration is essential for solving the complex challenges of our time.